\documentclass{PoS}
\title{
Calculation of $\rho$ meson decay width 
from the PACS-CS configurations
}
\ShortTitle{
Calculation of $\rho$ meson decay width 
from the PACS-CS configurations
}
\author{
PACS-CS collaboration:
}
\author{
S.~Aoki        ${}^{a,b}$,
K.-I.~Ishikawa ${}^{c  }$,
\speaker{N.~Ishizuka}${}^{a,b}$\thanks{E-mail : ishizuka@ccs.tsukuba.ac.jp},
K.~Kanaya      ${}^{a  }$,
Y.~Kuramashi   ${}^{a,b}$,
Y.~Namekawa    ${}^{b  }$,
M.~Okawa       ${}^{c  }$,
Y.~Taniguchi   ${}^{a,b}$,
A.~Ukawa       ${}^{a,b}$,
N.~Ukita       ${}^{b  }$,
T.~Yamazaki    ${}^{b  }$,
T.~Yoshi\'e    ${}^{a,b}$
\\ \\ \\
\llap{$^a$}
Graduate School of Pure and Applied Sciences,
University of Tsukuba, Tsukuba 305-8571, Japan.
\\
\llap{$^b$}
Center for Computational Sciences,
University of Tsukuba, Tsukuba 305-8577, Japan.
\\
\llap{$^c$}
Department of Physics, Hiroshima University,
Higashi-Hiroshima 739-8526, Japan.
}
\FullConference{
The XXVIII International Symposium on Lattice Field Theory, \\
Lattice2010         \\
June 14-19, 2010    \\
Villasimius, Italy
}
\abstract{
We present preliminary results on the $\rho$ meson decay width
from $N_f=2+1$ full QCD configurations 
generated by PACS-CS Collaboration.
The decay width
is estimated from the $P$-wave scattering phase shift
for the isospin $I=1$ two-pion system.
The finite size formula
presented by L\"uscher in the center of mass frame and 
its extension to non-zero total momentum frame
by Rummukainen and Gottlieb
are employed for the calculations of the phase shift.
Our calculations are carried out 
at $m_\pi=410\ {\rm MeV}$ ($m_\pi/m_\rho=0.46$)
and $a=0.091\ {\rm fm}$
on a $32^3\times 64$ ($La=2.9 {\rm fm}$) lattice.
}
\begin{document}
%
%
\section{ Introduction }
Study of the $\rho$ meson decay
is a significant step for understanding
the dynamical aspects of hadron interactions with lattice QCD.
In the early stage of studies toward this direction
the transition amplitude $\langle \rho | \pi\pi \rangle$
extracted from the time behavior of the correlation function
$\langle \pi(t)\pi(t) \rho(0) \rangle$
has been used to estimate the decay width,
assuming that the hadron interaction
is small~\cite{rhd:TAMP:GMTW,rhd:TAMP:LD,rhd:TAMP:MM,rhd:TAMP:JMMU}.

A more realistic approach is an estimation of the decay width 
from the $P$-wave scattering phase shift
for the isospin $I=1$ two-pion system.
The finite size formula
presented by L\"uscher in the center of mass frame~\cite{Lfm:L} or
its extension to non-zero total momentum frame
by Rummukainen and Gottlieb~\cite{rhd:SCPH:CP-PACS}
is employed for an estimation of the phase shift.
The first study of this approach
was carried by CP-PACS Collaboration
using $N_f=2$ full QCD configurations 
($m_\pi=330\ {\rm MeV}$, $a=0.21\ {\rm fm}$, 
$La=2.5\ {\rm fm}$)~\cite{rhd:SCPH:CP-PACS}.
After this work,
two studies were reported
with $N_f=2$ configurations on finer lattices,
one by QCD-SF Collaboration ($m_\pi=240-810\ {\rm MeV}$ , 
$a=0.072-0.084\ {\rm fm}$)~\cite{rhd:SCPH:QCDSF}
and the other by ETMC Collaboration
($m_\pi=391\ {\rm MeV}$, $a=0.086\ {\rm fm}$, 
$La=2.1\ {\rm fm}$)~\cite{rhd:SCPH:ETMC_1}.

In the present work 
we extend these studies by employing 
$N_f=2+1$ full QCD configurations
and working on a larger lattice volume.
Our calculations are carried out on a subset of 
configurations previously generated by PACS-CS Collaboration
with the Iwasaki gauge action and nonperturbatively
$O(a)$-improved Wilson fermion action at $\beta=1.9$.  
on a $32^3\times 64$ lattice~\cite{conf:PACS-CS}.
The subset corresponds to the hopping parameters
$\kappa_{ud}=0.13754$ for the up and down quark and 
$\kappa_{s }=0.13640$ for the strange quark. 
The parameters determined from the spectrum analysis 
for this subset are 
$m_\pi=410\ {\rm MeV}$ ($m_\pi/m_\rho=0.46$), $a=0.091\ {\rm fm}$
and $La=2.91\ {\rm fm}$.
All calculations are carried out on
the PACS-CS computer at Center for Computational Sciences,
University of Tsukuba.
We note that ETMC~\cite{rhd:SCPH:ETMC_2}  
and BMW~\cite{rhd:SCPH:BMW} Collaborations 
reported their preliminary results at Lattice 2010.
%
%
\section{ Method }
We consider
the center of mass frame (CM)
and the non-zero total momentum frame (the moving frame (MF))
with the total momentum ${\bf p}=(2\pi/L){\bf e}_3$.
In these frames
the ground ($n=1$) and the first exited states ($n=2$)
with spin $J=1$ and isospin $I=1$, 
ignoring hadron interactions, are given by 
\begin{equation}
\begin{array}{lc cc lc lc lc l}
\mbox{frame}&\ & {\bf p} L/(2\pi) &\ & {\rm g} &\ & \Gamma
&\ \ & n=1         
&\ \ & n=2                                
\\
\mbox{CM}   &  & (0,0,0)          &  & O_h     &  & {\bf T}^{-}_1  
&& \mbox{\underbar{$\rho_{j}({\bf 0})$}}                                       
&& \pi({\bf e}_j)\pi(-{\bf e}_j)\ [ 1.3  ]  
\\
\mbox{MF}   &  & (0,0,1)          &  & D_{4h}  &  & {\bf E}^{-}    
&& \mbox{\underbar{$\rho_{1,2}(0,0,1)$}}                                                                  
&& \pi(1,0,1)\pi(-1, 0, 0) \ , \ 
   \pi(0,1,1)\pi( 0,-1, 0) \  [ 1.4  ]  
\\
\mbox{MF}   &  & (0,0,1)          &  & D_{4h}  &  & {\bf A}^{-}_2  
&& \mbox{\underbar{$\rho_{3}(0,0,1)$}}    
&& \mbox{\underbar{$\pi(0,0,1)\pi( 0, 0, 0) \  [ 1.02 ]$ }} \\
\end{array}
\ \ , 
\label{eq:energy_level}
\end{equation}
where ${\bf p}$ is the total momentum,
${\rm g}$ is the rotational group on the lattice and
$\Gamma$ is the irreducible representation of the group.
The vectors in parentheses after $\pi$ and $\rho$ refer 
to the momenta of the pion and the $\rho$ meson in units of $2\pi/L$.
The numbers in square brackets of the two-pion states are 
values of $\sqrt{s}/m_\rho$ on our full QCD configurations.
In the present work 
we calculate the scattering phase shifts 
of the states marked by under-bar in (\ref{eq:energy_level}).
The finite size formulas for these states 
are given in Refs.~\cite{Lfm:L,Lfm:RG}.

For the ${\bf T}^{-}_1$ and the ${\bf E}^{-}$ representations 
the energy of the ground state
is much smaller than that of the exited state
as one can see from the value of $\sqrt{s}/m_\rho$ in (\ref{eq:energy_level}).
Thus the energy of these states
can be extracted by a single exponential fit for 
the time correlation functions of the $\rho$ meson.
We use the local $\rho$ meson operator for the sink
and a smearing operator for the source
as discussed later.

For the ${\bf A}^{-}_2$ representation
the energy of the first exited state
is close to that of the ground state.
Thus we use the variational method~\cite{method_diag}
with a matrix of the time correlation function,
\begin{equation}
G(t) =   \left(
  \begin{array}{ll}
     \     \langle 0 | \ \Omega^\dagger(t) \ \overline{\Omega}(t_s) \ | 0 \rangle
&    \quad \langle 0 | \ \Omega^\dagger(t) \ \overline{\rho_3}(t_s) \ | 0 \rangle  \\
     \     \langle 0 | \ \rho_3^\dagger(t) \ \overline{\Omega}(t_s) \ | 0 \rangle
&    \quad \langle 0 | \ \rho_3^\dagger(t) \ \overline{\rho_3}(t_s) \ | 0 \rangle
  \end{array}
  \right)
\ .
\label{eq:G_def}
\end{equation}
We extract the energy
by a single exponential fit for 
the two eigenvalues $\lambda_1 (t)$ and $\lambda_2 (t)$
of the matrix $M(t) = G(t) G^{-1}(t_R)$
with some reference time $t_R$,
assuming that the lower two states dominate 
the correlation function.
In (\ref{eq:G_def}),
$\rho_3(t)$ is the local operator
for the neutral $\rho$ meson at time $t$ with momentum
${\bf p}$ and the polarization vector parallel to the momentum.
$\Omega(t)$ is an operator
for the two pions
with the momentum ${\bf 0}$ and ${\bf p}=(2\pi/L){\bf e}_3$,
\begin{equation}
  \Omega(t) = \frac{1}{\sqrt{2}}
  \Bigl(
        \pi^{+}({\bf 0},t_1) \pi^{-}({\bf p},t)
      - \pi^{-}({\bf 0},t_1) \pi^{+}({\bf p},t)   
  \Bigl) \times {\rm e}^{ m_\pi \cdot ( t_1 - t ) }
\ ,
\label{eq:pp_op_sink}
\end{equation}
where 
$\pi({\bf p},t)$ is the local pion operator 
with momentum ${\bf p}$ at time $t$.
The times slice of the pion with zero momentum
is fixed at $t_1 >> t $,
and an exponential time factor is introduced,
so that the operator
has the same time behavior as that of the usual Heisenberg operator
for $t_1 >> t$,
{\it ie.}, 
$\langle 0 | \Omega^\dagger (t) = \langle 0 | \Omega^\dagger (0) {\rm exp}( - H \cdot t )$.

Two operators $\overline{\rho_3}(t_s)$ and $\overline{\Omega}(t_s)$
are used for the source in (\ref{eq:G_def}),
which are given by
\begin{eqnarray}
&&
\overline{\Omega}(t_s)
= \frac{1}{\sqrt{2}}
   \Bigl(   \pi^{+}({\bf 0},t_s) \pi^{-}({\bf p},t_s)
          - \pi^{-}({\bf 0},t_s) \pi^{+}({\bf p},t_s)   \Bigl)
\ , 
\label{eq:pp_op_source} 
\\
&&
\overline{\rho_3}(t_s) = 
  \sum_{{\bf z}\in \Gamma}
     \frac{1}{\sqrt{2}}
     \Bigl(   \overline{U}({\bf z},t_s) \gamma_3 U({\bf z},t_s)
            - \overline{D}({\bf z},t_s) \gamma_3 D({\bf z},t_s)   \Bigl)
   {\rm e}^{ i {\bf p}\cdot{\bf z} } 
\ .
\label{eq:rho_op_source} 
\end{eqnarray}
The operator
$U({\bf z},t_s)$ ($D({\bf z},t_s)$) 
is a smearing operator 
for the up (down) quark given by 
$
  U({\bf z},t_s) = \sum_{\bf_x} u( {\bf z} + {\bf x},t_s ) 
                   \cdot F(|{\bf x} - {\bf z}|)
$,
where $u({\bf x},t_s )$ is the quark operator 
at position ${\bf x}$ and time $t_s$.
We use the same smearing function
$F(x)$ as in Ref.~\cite{conf:PACS-CS}.
This operator is used 
after fixing gauge configurations to the Coulomb gauge.
In (\ref{eq:rho_op_source})
a summation over ${\bf z}$ is taken to reduce
contaminations from the states with different total momenta
and
$\Gamma = \{\ {\bf z}\ |\ {\bf z}=(L/2)\cdot(n_1,n_2,n_3) 
\ , \ n_j = \mbox{$0$ or $1$}\ \}$ is chosen
in the present work.
The smearing operator (\ref{eq:rho_op_source})
is also used to extract the energy
of the ground state for the ${\bf T}_1^-$
and the ${\bf E}^{-}$ representations
setting the momentum to ${\bf p}={\bf 0}$ and $(2\pi/L){\bf e}_3$.

The quark contractions of $G(t)$ in (\ref{eq:G_def}) are given by
\begin{equation}
\includegraphics[width=13.5cm]{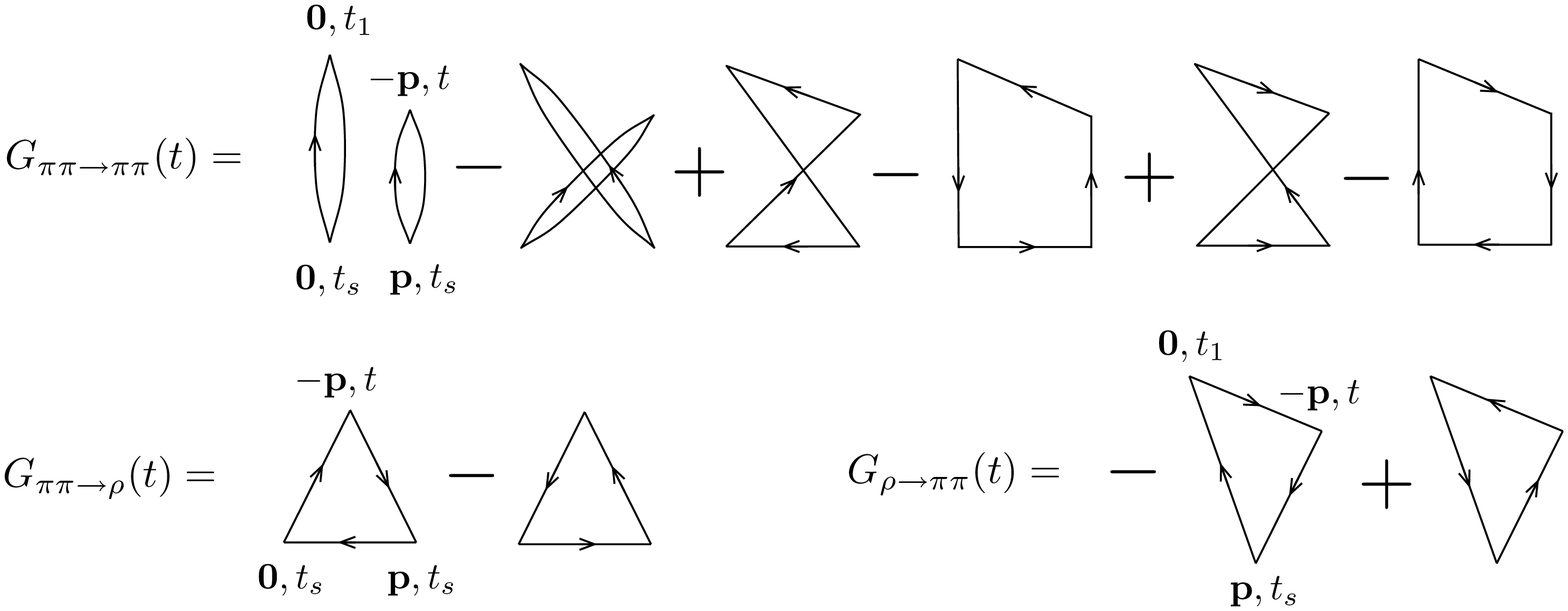}
\label{eq:quark_c}
\end{equation}
where the time runs upward in the diagrams.
The vertices refer to the pion or the $\rho$ meson operator 
with the momentum at the time specified in the diagrams.
The $\rho$ meson operators at $t_s$ are the smearing operators,
and the other is the local operator.

In order to calculate the quark contractions in (\ref{eq:quark_c}),
we use the source method and the stochastic noise method
as in the previous work by CP-PACS Collaboration~\cite{rhd:SCPH:CP-PACS}.
We introduce 
a $U(1)$ noise $\xi_j({\bf x})$ 
which satisfies
\begin{equation}
 \sum_{j=1}^{N_R}
 \xi_j^\dagger ({\bf x}) \xi_j ({\bf y})
  = \delta^3( {\bf x} - {\bf y} )
             \qquad \mbox{ for \ $N_R \to \infty$ }
\ ,
\label{eq:xi_prop}
\end{equation}
where $N_R$ is the number of noises
taken to be $10$ in the present work.
We calculate the following four types of quark propagators :
\begin{eqnarray}
&&
  Q_{AB}( {\bf x}, t | {\bf q}, t_s, \xi_j )
    = \sum_{\bf y} ( D^{-1} )_{AB}({\bf x}, t ; {\bf y}, t_s )
      \cdot \Bigl[ {\rm e}^{ i{\bf q}\cdot{\bf y}} \xi_j({\bf y}) \Bigr]
\ ,
\label{eq:QP_Q}
\\
&&
  W_{AB} ( {\bf x}, t | {\bf k}, t_a | {\bf q}, t_s, \xi_j )
  = \sum_{\bf y}
    \sum_{C}
       ( D^{-1} )_{AC}({\bf x}, t ; {\bf z}, t_a )
         \cdot
         \Bigl[ {\rm e}^{ i{\bf k}\cdot{\bf z}}
                      \gamma_5 \ Q( {\bf z}, t_a | {\bf q}, t_s, \xi_j )
         \Bigr]_{CB}
\ ,
\label{eq:QP_W}
\\
&&
  \overline{Q}_{AB}( {\bf x}, t | {\bf z}, t_s )
    = \sum_{\bf y} ( D^{-1} )_{AB}({\bf x}, t ; {\bf y}, t_s )
      \cdot \Bigl[ F(|{\bf y}-{\bf z}|) \Bigr]
\ ,
\label{eq:QP_Q_SM}
\\
&&
  \overline{W}_{AB} ( {\bf x}, t | {\bf k}, t_a | {\bf z}, t_s )
  = \sum_{\bf y}
    \sum_{C}
       ( D^{-1} )_{AC}({\bf x}, t ; {\bf y}, t_a )
         \cdot
         \Bigl[ {\rm e}^{ i{\bf k}\cdot{\bf y}}
                      \gamma_5 \ \overline{Q}( {\bf y}, t_a | {\bf z}, t_s )
         \Bigr]_{CB}
\ ,
\label{eq:QP_W_SM}  
\end{eqnarray} 
where $A$ and $B$ refer to color and spin indices,
and $F(x)$ is the smearing function.
The square bracket
is used as the source
for the inversion of the Dirac operator $D$.

The first term of the $\pi\pi\to\pi\pi$ component of (\ref{eq:quark_c})
can be calculated by introducing an another $U(1)$ noise $\eta_j({\bf x})$
having the same property as $\xi_j({\bf x})$ in (\ref{eq:xi_prop}),
\begin{equation}
\sum_{j}
\sum_{{\bf x}, {\bf y}} {\rm e}^{ - i {\bf p}\cdot {\bf y} }
\cdot
\Bigl\langle
       Q^\dagger ( {\bf x}, t_1 | {\bf 0}, t_s, \xi_j  ) \
       Q         ( {\bf x}, t_1 | {\bf 0}, t_s, \xi_j  ) \Bigr\rangle
\cdot {\rm e}^{ m_\pi \cdot ( t_1 - t ) }
\cdot
\Bigl\langle
       Q^\dagger ( {\bf y}, t   | {\bf p}, t_s, \eta_j ) \
       Q         ( {\bf y}, t   | {\bf p}, t_s, \eta_j ) \Bigr\rangle
\ ,
\label{eq:D1_quark_c}
\end{equation}
where the bracket means trace for the color and the spin indices.
The exponential time factor comes from the definition 
of the operator of the two pions in (\ref{eq:pp_op_sink}).
The second term
is given by exchanging the momentum and the time slice 
of the sink in (\ref{eq:D1_quark_c}).
The 3rd to 6th terms can be calculated by
\begin{eqnarray}
G^{\rm [3rd]}_{\pi\pi\to\pi\pi}(t) &=&
\sum_{j}
\sum_{\bf x}  {\rm e}^{ - i {\bf p}\cdot {\bf x} } \cdot
\Bigl\langle
\  W^\dagger ({\bf x},t| {\bf 0}, t_1 | {\bf 0}, t_s, \xi_j ) 
\  W         ({\bf x},t| {\bf 0}, t_s | {\bf p}, t_s, \xi_j ) 
\ \Bigr\rangle
\cdot {\rm e}^{ m_\pi \cdot ( t_1 - t ) }
\ ,
\cr
G^{\rm [4th]}_{\pi\pi\to\pi\pi}(t) &=&
\sum_{j}
\sum_{\bf x}  {\rm e}^{ - i {\bf p}\cdot {\bf x} } \cdot
\Bigl\langle
\  W^\dagger ({\bf x},t| {\bf 0}, t_1 | {\bf 0}, t_s, \xi_j ) 
\  W         ({\bf x},t| {\bf p}, t_s | {\bf 0}, t_s, \xi_j ) 
\  \Bigr\rangle
\cdot {\rm e}^{ m_\pi \cdot ( t_1 - t ) }
\ ,
\cr
G^{\rm [5th]}_{\pi\pi\to\pi\pi}(t) &=&
\sum_{j}
\sum_{\bf x}  {\rm e}^{ - i {\bf p}\cdot {\bf x} } \cdot
\Bigl\langle
\  W         ({\bf x},t| {\bf 0}, t_1 |  {\bf 0}, t_s, \xi_j )
\  W^\dagger ({\bf x},t| {\bf 0}, t_s | -{\bf p}, t_s, \xi_j )
\  \Bigr\rangle
\cdot {\rm e}^{ m_\pi \cdot ( t_1 - t ) }
\ ,
\cr
G^{\rm [6th]}_{\pi\pi\to\pi\pi}(t) &=&
\sum_{j}
\sum_{\bf x}  {\rm e}^{ - i {\bf p}\cdot {\bf x} } \cdot
\Bigl\langle
\  W         ({\bf x},t|  {\bf 0}, t_1 |  {\bf 0}, t_s, \xi_j )
\  W^\dagger ({\bf x},t| -{\bf p}, t_s |  {\bf 0}, t_s, \xi_j )
\  \Bigr\rangle
\cdot {\rm e}^{ m_\pi \cdot ( t_1 - t ) }
\ .
\label{eq:QC_B}
\end{eqnarray}
The two terms of $\pi\pi\to\rho$ of (\ref{eq:quark_c})
can be similarly obtained by
\begin{eqnarray}
G^{\rm [1st]}_{\pi\pi\to\rho}(t) &=&
\sum_{j}
\sum_{\bf x}  {\rm e}^{ - i {\bf p}\cdot {\bf x} } \cdot
\Bigl\langle
\  W^\dagger ({\bf x},t| - {\bf p}, t_s | {\bf 0}, t_s, \xi_j ) \ (\gamma_5 \gamma_3 ) 
\  Q         ({\bf x},t                 | {\bf 0}, t_s, \xi_j ) 
\  \Bigr\rangle
\ ,
\cr
G^{\rm [2nd]}_{\pi\pi\to\rho}(t) &=&
\sum_{j}
\sum_{\bf x}  {\rm e}^{ - i {\bf p}\cdot {\bf x} } \cdot
\Bigl\langle
\  Q^\dagger ({\bf x},t                | {\bf 0}, t_s, \xi_j ) \ (\gamma_5 \gamma_3 )
\  W         ({\bf x},t | {\bf p}, t_s | {\bf 0}, t_s, \xi_j ) 
\ \Bigr\rangle
\ .
\label{eq:QC_pprh}
\end{eqnarray}
We can calculate the two terms of $\rho\to\pi\pi$ of (\ref{eq:quark_c})
by
\begin{eqnarray}
G^{\rm [1st]}_{\rho\to\pi\pi}(t) &=&
\sum_{{\bf z}\in\Gamma} {\rm e}^{   i {\bf p}\cdot {\bf z} }
\sum_{\bf x}            {\rm e}^{ - i {\bf p}\cdot {\bf x} } \cdot
\Bigl\langle
\  \overline{Q}^\dagger ( {\bf x}, t                | {\bf z}, t_s )
\  \overline{W}         ( {\bf x}, t | {\bf 0}, t_1 | {\bf z}, t_s ) \ (\gamma_3 \gamma_5 )
\  \Bigr\rangle
\cdot {\rm e}^{ m_\pi \cdot ( t_1 - t ) }
\ ,
\cr
G^{\rm [2nd]}_{\rho\to\pi\pi}(t) &=&
\sum_{{\bf z}\in\Gamma} {\rm e}^{   i {\bf p}\cdot {\bf z} }
\sum_{\bf x}            {\rm e}^{ - i {\bf p}\cdot {\bf x} } \cdot
\Bigl\langle
\  \overline{W}^\dagger ( {\bf x}, t | {\bf 0}, t_1 | {\bf z}, t_s )
\  \overline{Q}         ( {\bf x}, t                | {\bf z}, t_s ) \ (\gamma_3 \gamma_5 )
\  \Bigr\rangle
\cdot {\rm e}^{ m_\pi \cdot ( t_1 - t ) }
\ .
\label{eq:QC_rhpp}
\end{eqnarray}
The $\rho\to\rho$ component is given 
by the $\overline{Q}$-type propagators as the usual two point function.

The quark propagators are 
calculated with the Dirichlet boundary condition
imposed in the time direction
and the source operators are set at $t_s=12$
to avoid effects from the temporal boundary.
We put the zero momentum pion
introduced in (\ref{eq:pp_op_sink}) at $t_1=42$.
We calculate the $Q$-type propagators (\ref{eq:QP_Q})
for four combinations of ${\bf q}$ and $U(1)$ noise : 
$
( {\bf q}, {\rm noise})=\{
( {\bf 0}, \xi  ) ,
( {\bf 0}, \eta ) ,
( {\bf p}, \xi  ) ,
(-{\bf p}, \xi  )
\}$.
The $W$-type propagators (\ref{eq:QP_W})
are calculated for five combinations
of ${\bf k}$, $t_a$ and ${\bf q}$ :
$
(  {\bf k}, t_a |  {\bf q} )=\{
(  {\bf p}, t_S |  {\bf 0} ) ,$ $
( -{\bf p}, t_S |  {\bf 0} ) ,$ $
(  {\bf 0}, t_S |  {\bf p} ) ,
(  {\bf 0}, t_S | -{\bf p} ) ,
(  {\bf 0}, t_1 |  {\bf 0} ) 
\}$, 
using the same $U(1)$ noise $\xi$ in common.
We calculate the $\overline{Q}$-type propagator (\ref{eq:QP_Q_SM})
and the $\overline{W}$-type propagator (\ref{eq:QP_W_SM})
with $({\bf k}, t_a)=({\bf 0},t_1)$ for the set ${\bf z}\in {\Gamma}$.
Thus we calculate 
$( 4 + 5 )\times 10 + ( 1 + 1 )\times 8 = 106$ quark propagators
for each configuration.
The total number of the configurations 
analyzed every $10$ trajectories is $440$.
We estimate the statistical error 
by the jackknife method with bins of $400$ trajectories.
%
%
\section{ Results }
\begin{figure}[t]
\includegraphics[width=7.5cm]{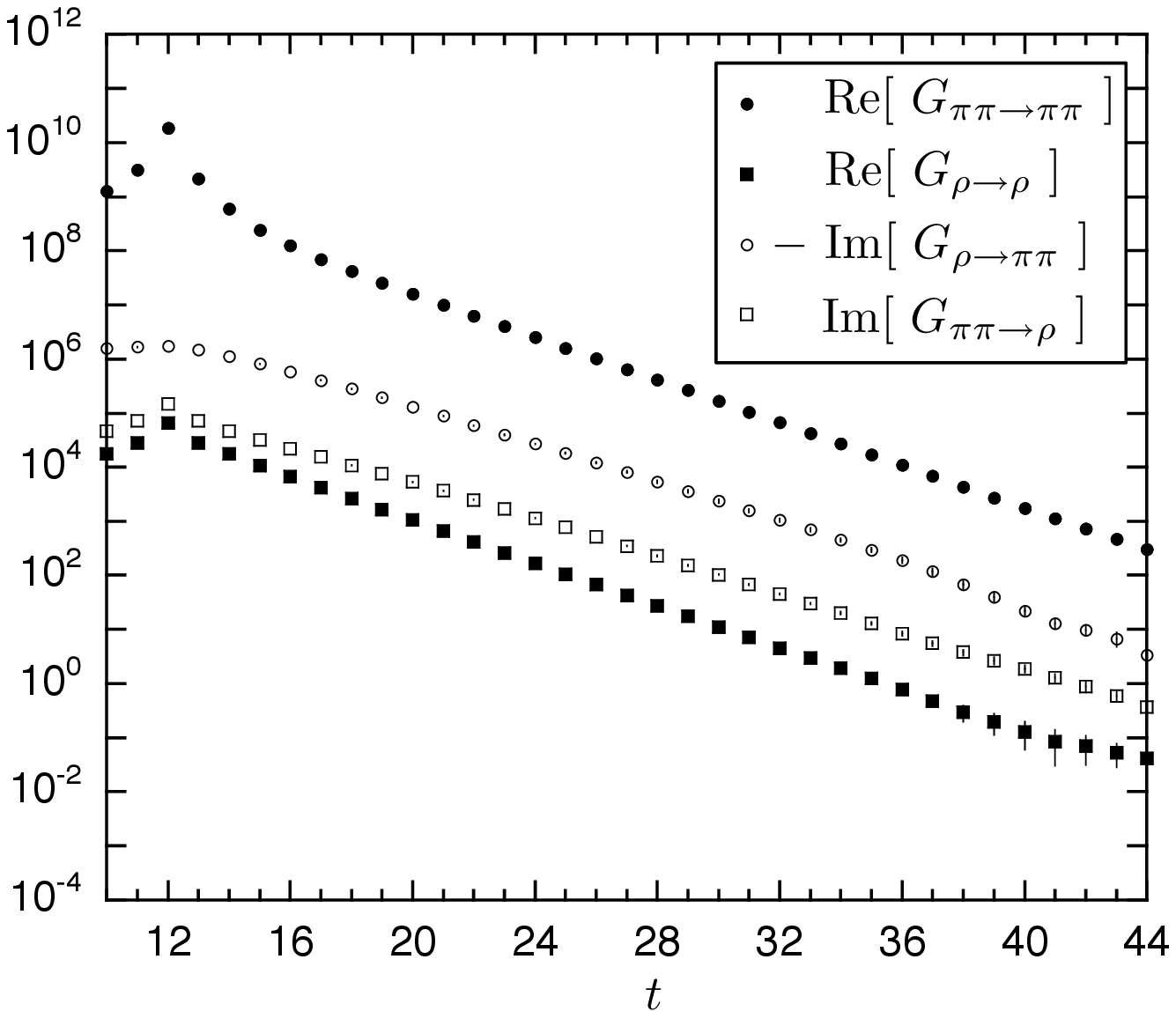} 
\includegraphics[width=7.5cm]{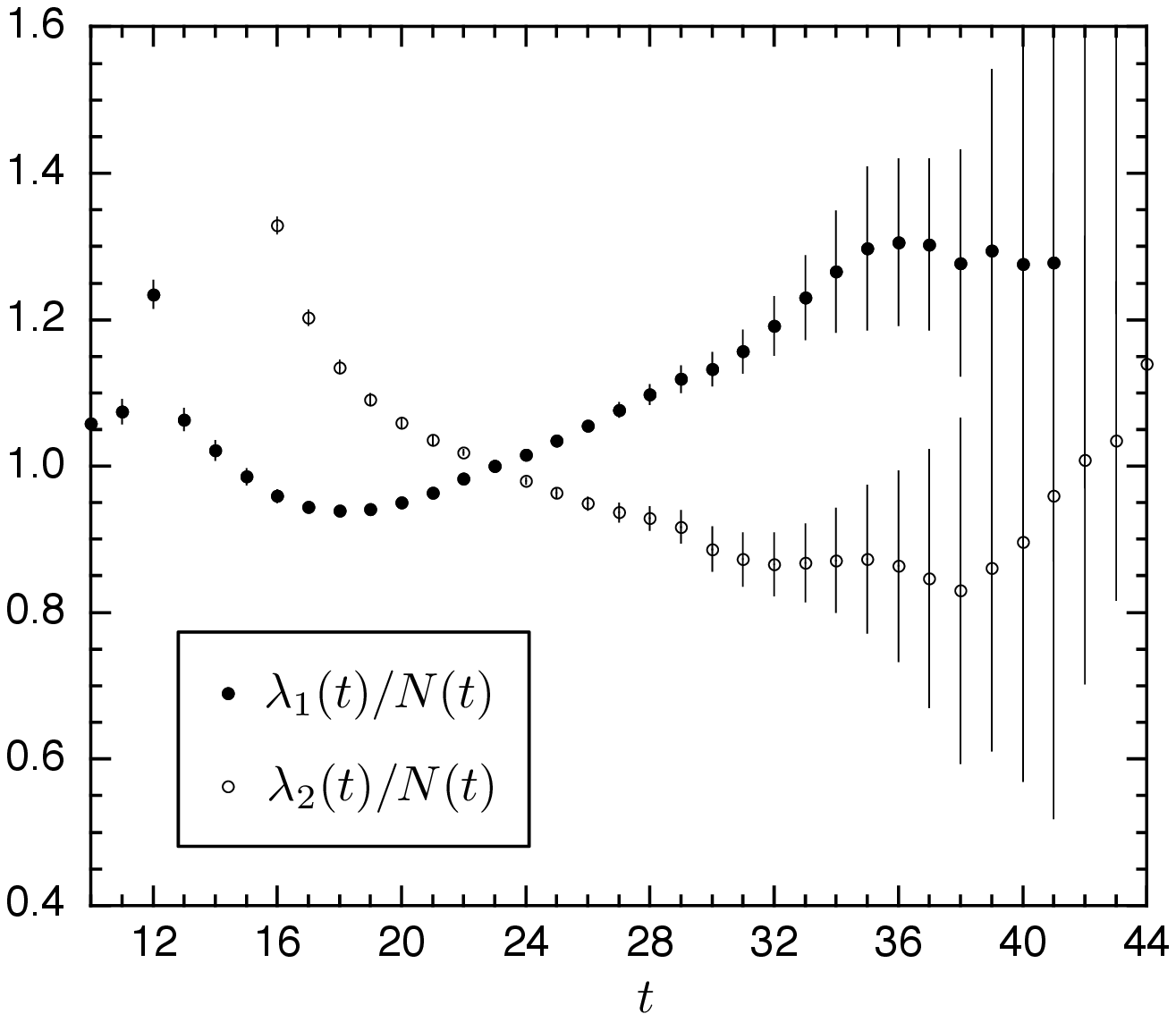}
\vspace{-0.6cm}
\caption{
Left panel : 
Real part of the diagonal components
($\pi\pi\to\pi\pi$ and $\rho\to\rho$)
and imaginary part of the off-diagonal components
($\pi\pi\to\rho$ and $\rho\to\pi\pi$)
of the time correlation function $G(t)$ ;
Right panel : 
Eigenvalues $\lambda_1(t)$ 
and $\lambda_2(t)$ normalized
by the correlation function of the two free pions $N(t)$.
}
\label{fig:G_t}
\end{figure}

In the left panel of Fig.~\ref{fig:G_t}
we show the real  part of the diagonal     components ($\pi\pi\to\pi\pi$ and $\rho\to\rho$)
and the imaginary part of the off-diagonal components ($\pi\pi\to\rho$, $\rho\to\pi\pi$)
of $G(t)$.
The other real or imaginary part of the components vanish
from $P$ and $CP$ symmetry.
We calculate 
the two eigenvalues $\lambda_n(t)$ ($n=1,2$)
for the matrix $M(t) = G(t) G^{-1}(t_R)$
with the reference time $t_R=23$.
In the right panel of Fig.~\ref{fig:G_t}
we plot the eigenvalues normalized
by the correlation function of the two free pions
$
N(t)=
\langle 0| \pi(-{\bf p},t) \pi({\bf p},t_s) |0\rangle
\langle 0| \pi( {\bf 0},t) \pi({\bf 0},t_s) |0\rangle
$.
Thus the slope of the figure corresponds to
the energy difference 
with respect to the energy of the two free pions.
We observe that
the energy difference
is  negative for $\lambda_1(t)$ 
and positive for $\lambda_2(t)$.
This means that the two-pion scatting phase shift
is positive for the ground state and negative for the first exited state
of the ${\bf A}_{2}^{-1}$ representation.
%

\begin{figure}[t]
\begin{center}
\includegraphics[width=9.0cm]{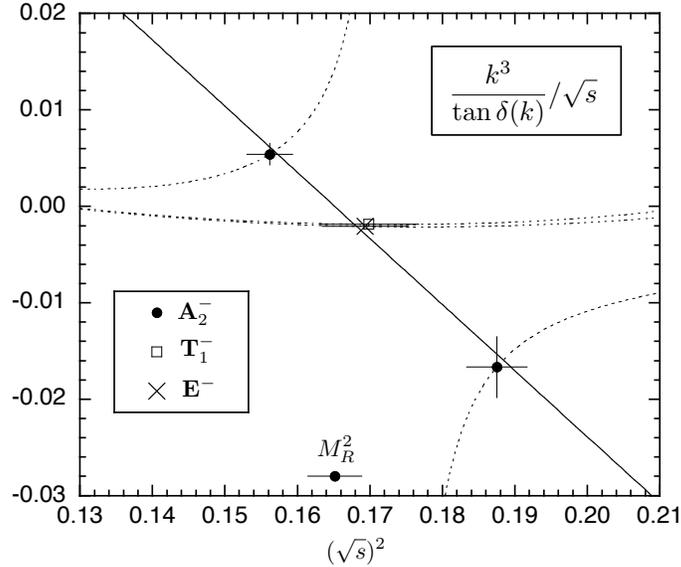}
\end{center}
\vspace{-0.6cm}
\caption{
$(k^3 / \tan\delta(k) )/\sqrt{s}$
as function of $(\sqrt{s})^2$ in unit of the lattice cutoff.
Dotted lines refer to the finite size formulas for each representation.
}
\label{fig:k2_AMP_SS}
\end{figure}

We extract the energies for both states
by a single exponential fitting of the eigenvalues
$\lambda_n(t)$ ($n=1,2$) for the time range $t=24-33$.
For the ${\bf T}^{-}_1$ and the ${\bf E}^{-}$ representations 
the energy of the ground state is obtained
from the time correlation functions of the $\rho$ meson
as explained before.
Converting the energies on each frame to the invariant masses $\sqrt{s}$
and substituting them into the finite size formulas,
we obtain the scattering phase shifts
plotted in Fig.~\ref{fig:k2_AMP_SS}.
In the figure 
we show $( k^3 / \tan\delta(k) ) / \sqrt{s}$
as function of $(\sqrt{s})^2$ in unit of the lattice cutoff,
where $k=\sqrt{ s/4 - m_\pi^2 }$
is the scattering momentum.
%

In order to estimate the $\rho$ meson decay width 
at the physical quark mass
we parametrize the scattering phase shift
with the effective $\rho\to\pi\pi$ coupling constant $g_{\rho\pi\pi}$
by 
\begin{equation}
\frac{k^3}{\tan\delta(k)}/\sqrt{s}
 = \frac{ 6\pi }{ g_{\rho\pi\pi}^2 } \cdot
   ( M_R^2 - s )
\ ,
\label{eq:k2_AMP_SS_fit}
\end{equation}
where $M_R$ is the resonance mass.
The coupling $g_{\rho\pi\pi}$
generally depends on the quark mass, 
but our present data
at a single quark mass do not provide this information.
Here we assume that the mass dependence is weak
and try to estimate $g_{\rho\pi\pi}$ and $M_R$
by fitting our results with (\ref{eq:k2_AMP_SS_fit}).
We estimate the $\rho$ meson decay width at the physical quark mass
from the formula
$\Gamma_\rho
  = g_{\rho\pi\pi}^2 \times 4.128 \ \ {\rm MeV}$.
Our results of the fitting are given by
\begin{eqnarray}
  a M_R          = 0.4064 \pm 0.046    \,\, , \quad
  g_{\rho\pi\pi} = 5.24   \pm 0.51     \,\, , \quad 
  \Gamma_\rho    = 113    \pm 22       \ \ {\rm MeV}
\ .
\label{eq:FinalR}
\end{eqnarray}
In Fig.~\ref{fig:k2_AMP_SS}
we indicate 
the result for the resonance mass $M_R$ and 
draw the  fitting curve by a solid line.

Our result of the $\rho$ meson decay width $\Gamma_\rho$ 
at the physical quark mass
is smaller than 
the experimental value ($150\ {\rm MeV}$).
Possible reasons of the discrepancy 
are the mass dependence of the effective coupling constant $g_{\rho\pi\pi}$
and the finite lattice spacing effects.
We will study these issues in next investigations.
%
%
\section{ Summary }
In the present work 
we estimated the $\rho$ meson decay width  
from the $N_f=2+1$ full QCD configurations 
generated by PACS-CS Collaboration.
The decay width
is estimated from the $P$-wave scattering phase shift
for $I=1$ two-pion system.
We used the effective $\rho\to\pi\pi$ coupling constant
$g_{\rho\pi\pi}$ to 
extrapolate from our simulation point $m_\pi = 410\ {\rm MeV}$ 
to the physical point $m_\pi = 135\ {\rm MeV}$,
assuming that $g_{\rho\pi\pi}$ does not depend on the quark mass.
The decay width may be estimated directly
from the energy dependence of the phase shift data
assuming the Breit-Wigner resonance formula,
if the simulations are made close to the physical quark mass
and we have data at several values of energy near the resonance mass.
We leave these issues to studies in the future.

This work is supported in part by Grants-in-Aid of the Ministry of Education
(Nos.
%
20340047, 20105001, 20105003
%
, 20740139
%
, 20540248
%
, 21340049
%
, 22244018, 20105002
%
, 22105501, 22740138
%
, 10143538
%
, 21105501
%
20105005 ).
The numerical calculations have been carried out
on PACS-CS at Center for Computational Sciences, University of Tsukuba.
%
%

%
%
\end{document}